\documentclass[lettersize,journal]{IEEEtran}

\usepackage{amsmath,amsfonts}
\usepackage{array}
\usepackage[caption=false,font=normalsize,labelfont=sf,textfont=sf]{subfig}
\usepackage{stfloats}
\usepackage{url}
\usepackage{verbatim}
\usepackage{graphicx}
\usepackage{cite}
\usepackage{amsmath,amssymb,amsfonts}
\usepackage{algorithm}
\usepackage{algpseudocode}
\usepackage{lipsum}
\usepackage{algorithmicx}
\usepackage{algpseudocode}
\usepackage{textcomp}
\usepackage{subcaption}
\usepackage{xcolor}
\usepackage{comment}
\usepackage{float} % for the 'H' specifier
\usepackage{hyperref} % This package adds hyperlink capabilities
\usepackage{listings}
\usepackage{xcolor}
\usepackage{adjustbox}
\usepackage{changepage}
\usepackage{booktabs}

\begin{document}

\title{LMDG: Advancing Lateral Movement Detection Through High-Fidelity Dataset
Generation\\
}
% 1\textsuperscript{st}

\author{Anas Mabrouk, Mohamed Hatem, Mohammad Mamun,~\IEEEmembership{Senior Member,~IEEE}, Sherif Saad,~\IEEEmembership{Senior Member,~IEEE}
        % <-this % stops a space

\thanks{This paper is based on the author's MSc thesis; for more details, see~\cite{mabrouk2024thesis}.\\

This work has been submitted to the IEEE for possible publication. Copyright may be transferred without notice, after which this version may no longer be accessible.}

% \thanks{This work has been submitted to the IEEE for possible publication. Copyright may be transferred without notice, after which this version may no longer be accessible.}

}

% \author{\IEEEauthorblockN{Anas Mabrouk and Sherif Saad}
% \IEEEauthorblockA{\textit{School of Computer Science} \\
% \textit{University of Windsor}\\
% Windsor, Canada \\
% \{mabrouka, shsaad\}@uwindsor.ca}
% \and

% \IEEEauthorblockN{Mohamed Hatem}
% \IEEEauthorblockA{\textit{Faculty of Information Engineering} \\
% \textit{German University Cairo}\\
% Cairo, Egypt \\
% mohamed.hatem@student.guc.edu.eg}

% \and
 
% \IEEEauthorblockN{Mohammad Mamun}
% \IEEEauthorblockA{\textit{ National Research Council Canada} \\
% % \textit{}\\
% Fredericton, Canada \\
% mohammad.mamun@nrc-cnrc.gc.ca}

% % \and

% % \IEEEauthorblockN{Sherif Saad}
% % \IEEEauthorblockA{\textit{School of Computer Science} \\
% % \textit{University of Windsor}\\
% % Windsor, Canada \\
% % shsaad@uwindsor.ca}

% }

\maketitle

\begin{abstract}

Lateral Movement (LM) attacks continue to pose a significant threat to enterprise security, enabling adversaries to stealthily compromise critical assets. However, the development and evaluation of LM detection systems are impeded by the absence of realistic, well-labeled datasets. To address this gap, we propose \textit{LMDG}, a reproducible and extensible framework for generating high-fidelity LM datasets. \textit{LMDG} automates benign activity generation, multi-stage attack execution, and comprehensive labeling of system and network logs, dramatically reducing manual effort and enabling scalable dataset creation. A central contribution of \textit{LMDG} is \textit{Process Tree Labeling}, a novel agent-based technique that traces all malicious activity back to its origin with high precision. Unlike prior methods such as \textit{Injection Timing} or \textit{Behavioral Profiling}, \textit{Process Tree Labeling} enables accurate, step-wise labeling of malicious log entries, correlating each with specific attack step and \textit{MITRE ATT\&CK} TTPs. To our knowledge, this is the first approach to support fine-grained labeling of multi-step attacks, providing critical context for detection models such as attack path reconstruction. We used \textit{LMDG} to generate a 25-day dataset within a 25-VM enterprise environment containing 22 user accounts. The dataset includes 944 GB of host and network logs and embeds 35 multi-stage LM attacks, with malicious events comprising less than 1\% of total activity—reflecting realistic benign-to-malicious ratio for evaluating detection systems. \textit{LMDG} generated datasets improves upon existing ones by offering diverse LM attacks, up-to-date attack patterns, longer attacks timeframes, comprehensive data sources, realistic network architectures, and more accurate labeling.

% This document is a model and instructions for \LaTeX.
% This and the IEEEtran.cls file define the components of your paper [title, text, heads, etc.]. *CRITICAL: Do Not Use Symbols, Special Characters, Footnotes, 
% or Math in Paper Title or Abstract.
\end{abstract}


\section{Introduction}
\label{sec:introduction}

\IEEEPARstart{A}{dvanced} Persistent Threats (APTs) represent a sophisticated category of cyberattacks characterized by their prolonged and stealthy presence within a targeted computer system or network, aimed at ultimately exfiltrating sensitive data or causing significant harm \cite{APTSurvey1, APTSurvey5, APTSurvey4}. APTs employ a diverse array of techniques and tactics meticulously crafted to circumvent the defensive mechanisms of the victim's security infrastructure \cite{mitre}. Among the array of sophisticated techniques employed by advanced threat actors, the concept of LM has emerged as a critical strategy for adversaries seeking to maneuver within compromised network environments. As elucidated by the exposition in \cite{mitre}, LM embodies an array of methodologies engaged by malevolent entities to infiltrate and orchestrate control over remote network systems. The attainment of their intended goals is frequently characterized by the imperative act of pivoting across an assortment of interconnected systems and accounts. Corresponding definitions mirroring this conception of LM are also extant within the literature, as expounded upon in \cite{hopper}, \cite{PivotingDetectiondataset}, \cite{latte}, and \cite{Graph-BasedImpactMetricForMitigatingLateralMovementCyberAttacks}, delineating the concept as the orchestrated movement of an attacker from a primary host to successive nodes within a compromised network, culminating in the pursuit of a valuable target.

LM-based attacks are becoming a growing threat to large private and government networks, frequently causing information exfiltration and service disruptions \cite{AnUnsupervisedMulti-DetectorApproachforIdentifyingMaliciousLateralMovement}. Analyzing various APT campaigns reveals that nearly all employ LM   to navigate networks. The purpose of LM   is to transition from one system to another, infiltrating additional resources and gaining higher privileges. This process enables attackers to discover and collect valuable data, expand their control over the targeted organization, and maintain long-term access to the compromised IT infrastructure \cite{Advancedpersistentthreats:Behindthescenes, APTSurvey1, APTSurvey5, AdvancedPersistentThreats(APT):EvolutionAnatomyAttributionAndCountermeasures}. Since LM   is a crucial phase in an APT attack, early detection is vital to minimize losses and prevent attackers from gaining further access to the network \cite{UncoveringLateralMovementUsingAuthenticationLogs}.

Detecting LM   attacks poses a significant challenge, primarily due to several factors; firstly, the prolonged duration of these attacks, which can extend over months, significantly complicates their detection. Additionally, the sheer volume of enterprise traffic provides adversaries ample opportunities to blend in and seamlessly remain undetected amidst regular network activity. Various tactics and techniques exist for executing LM attacks, often leaving traces within network and system logs \cite{mitre}. Attackers can effectively evade detection mechanisms by leveraging legitimate authentication credentials, system tools, and other evasion techniques. Furthermore, the prevalence of false security alerts further adds to the difficulty of distinguishing genuine threats from benign anomalies. Moreover, the incorporation of zero-day exploits or novel malware variants as part of these attacks further amplifies the complexity of detection \cite{UncoveringLateralMovementUsingAuthenticationLogs, Unicorn,DetectingLateralMovementInEnterpriseComputerNetworks,RDP-basedLateralMovementdetection,DetectionThreatPrioritizationOfPivotingAttacks}.

Current research endeavors for LM   detection rely mainly on machine learning \cite{OnTheDetectionOfLateralMovementThroughSupervisedMachineLearningAndAnOpen-SourceToolTocreateTurnkeyDatasetsFromSysmonLogs, AnUnsupervisedMulti-DetectorApproachforIdentifyingMaliciousLateralMovement, MLTracer:MaliciousLoginsDetectionSystemViaGraphNeuralNetwork, HetGLM, AnUnsupervisedApproachForDetectingLateralMovementLoginsBasedOnKnowledgeGraph}. The machine learning paradigm depends heavily on datasets to train and evaluate detection models, and the quality of these datasets directly impacts model performance and evaluation accuracy. Without high-quality training data, models can exhibit performance discrepancies, reducing accuracy and increasing false positives \cite{BringingThePeopleBackIn:ContestingBenchmarkMachineLearningDatasets, ArePublicIntrusionDatasetsFitForPurposeCharacterisingTheStateOfTheArtInIntrusionEventDatasets}. A growing body of literature explores the evidence supporting that neglecting the fundamental importance of data has led to inaccuracies and bias in ML models \cite{DataPerf}. For instance, researchers in \cite{QualityIn/QualityOut} demonstrated that even minor modifications to a benchmark dataset significantly impact model performance more than the specific machine learning technique. Therefore, better data quality is essential to improve generalization and avoid bias in machine learning models \cite{DataExcellenceForAI:WhyShouldYouCare?, EveryoneWantsToDoTheModelWorkNotTheDataWork}.

 Most cybersecurity datasets suffer from quality issues, particularly those containing LM   attacks. Common data quality problems include noisy labels, insufficient labeling, class imbalance, limited diversity of attack patterns, outdated attack types, simplistic synthetic generation environments, and short generation periods (see section \ref{sec:related_work}). Additionally, many existing datasets either lack instances of LM   attacks altogether or contain only a limited number of such instances\cite{DataCurationAndQualityEvaluationForMachineLearning-BasedCyberIntrusionDetection, ArePublicIntrusionDatasetsFitForPurposeCharacterisingTheStateOfTheArtInIntrusionEventDatasets, DatasetsAreNotEnough:ChallengesInLabelingNetworkTraffic, BridgingTheGapToReal-worldForNetworkIntrusionDetectionSystemsWithData-CentricApproach}. Consequently, developing a comprehensive dataset, or ideally a framework, that addresses these challenges and others is essential for advancing research in LM   detection.

 To this end, our paper introduces a framework called LMDG (LM   Datasets Generator), which addresses most of the issues discussed in section \ref{sec:related_work}. Our contributions can be summarized as follows:

\begin{itemize}
  \item We analyzed existing cybersecurity datasets for LM attack properties, including techniques, time frames, movement hops, data sources, labeling methods, and testbed designs. We also reviewed frameworks for creating LM datasets \ref{sec:related_work}, making this the first study focused on LM dataset evaluation.\\

  \item Creating a benchmark dataset focused on LM   attacks that address many of the existing issues in current LM datasets and conducting a qualitative analysis of it \ref{sec:dataset} \ref{sec:qualitative_analysis}. This dataset will be valuable for the research community in training and evaluating LM detection models.\\

  \item We developed the LMDG framework to generate reproducible LM/APT datasets (see Section \ref{sec:lmdg}). It automates benign data generation \ref{subsec:BDE}, attack execution \ref{subsec:attack_engine}, and labeling. Automatic labeling is challenging in LM datasets due to malicious activities by benign hosts. Existing techniques include Injection Timing, Behavior Profiles, and Network Security Tools \cite{DatasetsAreNotEnough:ChallengesInLabelingNetworkTraffic, Human-guidedAuto-LabelingForNetworkTrafficData:TheGELMapproach}. We introduce a new method, \textbf{\textit{process tree labeling}}, which we argue is the most accurate \ref{subsec:labelling_engine}.

\end{itemize}

\section{LMDG Framework}
\label{sec:lmdg}

\subsection{Overview}

The LMDG framework uses VirtualBox to simulate organizational networks for LM detection research. While synthetic datasets offer controlled conditions and flexible attack simulations, they may lack realism. To address this, LMDG employs advanced virtualization techniques to enhance dataset fidelity. VirtualBox's networking options—NAT, Bridged, Internal, and Host-Only networks—allow for flexible and realistic network simulations. These configurations help emulate diverse organizational topologies, improving the practicality of the generated datasets for cybersecurity research.

Active Directory (AD) is a widely used directory service for managing IT profiles, enabling authentication, authorization, and resource management \cite{ActiveDirectoryAttacks—StepsTypesAndSignatures, Ransomware:AnalysingTheImpactOnWindowsActiveDirectoryDomainServices}. The recent CrowdStrike incident highlighted the critical reliance on Windows systems, causing major disruptions and financial losses \cite{CrowdStrike_2024, kerner2024crowdstrike}. The LMDG framework integrates Windows Domains and AD to create realistic network environments for cybersecurity research, improving the accuracy of simulated datasets for studying advanced threats like APTs.

The LMDG framework uses Wireshark for capturing detailed network traffic traces and employs a persistent logging service on all hosts and gateways to ensure data continuity. It also leverages Windows Event Logs for comprehensive system and security event records. This combination ensures high-fidelity data collection, crucial for developing effective cybersecurity detection models.

\subsection{Testbed Infrastructure}
\label{subsec:testbed}

\begin{figure*}[h]
    \centering
    \includegraphics[width=\textwidth]{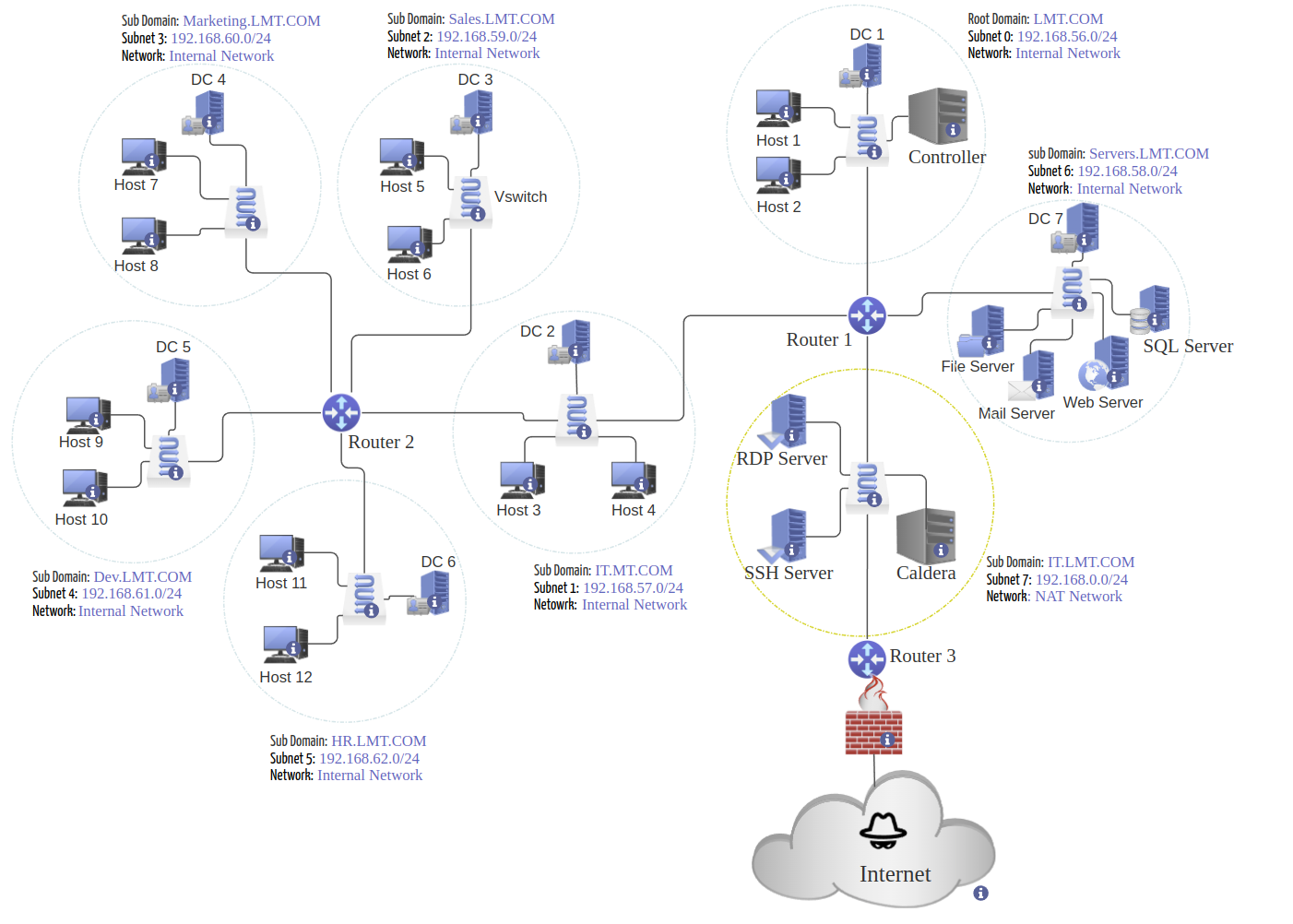} % full width
    \caption{The Testbed architecture used to generate LMDG dataset.}
    \label{fig:network_topology}
\end{figure*}

The network, shown in Figure \ref{fig:network_topology}, simulates a small company with five departments, each in its own subnet with a dedicated Windows domain. For example, the Sales department resides in \textit{sales.lmt.com} on subnet \textit{192.168.59.0/24} with domain controller \textit{DC 3}. Additional subnets include the root domain \textit{lmt.com}, company servers, and a DMZ (\textit{192.168.0.0/24}) within the IT domain. Routers connect these segments, and the structure can be adjusted as needed.

The setup uses VirtualBox with internal networks for subnets (isolated from external traffic) and a NAT network for the DMZ to enable controlled external communication. Hosts run Windows 10/11, while servers use Windows Server 2022 to reflect modern enterprise environments.

This topology improves realism by mirroring real-world enterprise networks with multiple subnets, dedicated domains, and a DMZ for public-facing services. These features enhance dataset fidelity, making it valuable for cybersecurity research.

\subsection{Benign Data Engine (BDE)}
\label{subsec:BDE}

\begin{figure*}[h]
    \centering
    \includegraphics[width=\textwidth]{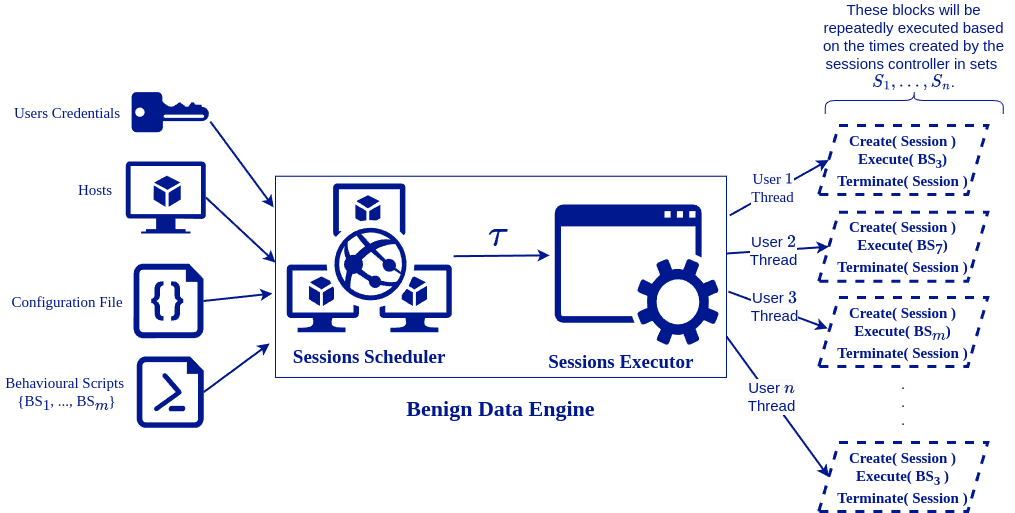}
    \caption{Benign Data Engine (BDE) overview.}
    \label{fig:BDE}
\end{figure*}

In the context of the LMDG framework, the Benign Data Engine (BDE) is tasked with generating normal network behavior, effectively simulating employee activities. Figure \ref{fig:BDE} provides an overview of the BDE engine, which comprises two primary components: the \textbf{Sessions Scheduler} \ref{ssubsec:sessions_scheduler} and the \textbf{Sessions Executor} \ref{ssubsec:sessions_exec}. The engine operates based on four key inputs:

\begin{itemize}
    \item \textbf{User Credentials and Hosts:} This includes the credentials of employees and the specific hosts (workstations or devices) they use within the network.

    \item \textbf{Sessions Scheduler Configuration File:} This file defines the parameters for the Sessions Scheduler, dictating how it should generate and manage sessions timing for each user or employee.
    
    \item \textbf{Behavioral Scripts:} These scripts detail the activities of employees, it can operate on both individual level and a departmental level, such as those specific to the IT department. They encapsulate routine tasks and behaviors expected in a typical workday.
\end{itemize}

The Sessions Scheduler orchestrates generating session behaviors (i.e., login and logout times), ensuring the simulated activities align with realistic standard user behavior patterns. Concurrently, the Sessions Executor enables the efficient simulation of multiple user sessions, reflecting the concurrent activities of various employees within the network. This design enhances the generated data's realism and ensures scalability and performance in simulating complex network environments.

\subsubsection{\textbf{Sessions Scheduler}}
\label{ssubsec:sessions_scheduler}

The role of the Sessions Scheduler is to generate a list of tuples \( \mathcal{S}_i = [(t_1, t_2), (t_3, t_4), \ldots, (t_{2k_i-1}, t_{2k_i})] \) for each employee \( i \), representing their session behavior. Each tuple \( (t_{2j-1}, t_{2j}) \) denotes the login and logout times of a session, where \( t_{2j-1} \) is the login time and \( t_{2j} \) is the logout time. The duration of a session is given by \( t_{2j} - t_{2j-1} \). The final output of the Sessions Scheduler is a list of lists \( \mathcal{T} = [\mathcal{S}_1, \mathcal{S}_2, \ldots, \mathcal{S}_n] \), where \( \mathcal{T} \) encapsulates the session behaviors for all \( n \) employees in the network. Each \( \mathcal{S}_i \) in \( \mathcal{T} \) provides a detailed account of an individual employee's login and logout activities throughout the day.

The process by which the Sessions Scheduler generates the lists \( \mathcal{S}_i \) for an employee \( i \) is outlined as follows. Initially, the Sessions Scheduler determines whether employee \( i \) is absent based on probability values specified in the configuration file (third input, Figure \ref{fig:BDE}) defined by the dataset creators. For instance, the dataset creators can define a probability interval \([p_1, p_2]\), where \( 0 \leq p_1, p_2 \leq 1 \). The Sessions Scheduler then selects a random value from this interval to represent the probability of employee \( i \) being absent on a given day. This approach ensures that each employee \( i \) has a distinct probability of being absent. If employee \( i \) is absent, then the list \( \mathcal{S}_i = \varnothing \). Additionally, there is a separate probability interval \([p'_1, p'_2]\) for determining absences during weekends, which typically corresponds to a higher probability.

If employee \(i\) is not absent, the Sessions Scheduler will proceed to generate the list \( \mathcal{S}_i \). Initially, it determines the starting time (first login) for employee \(i\). To facilitate this, the dataset creators define four \textit{time intervals} representing various starting times: abnormally early, abnormally late, late, and on time. These intervals are denoted as \([t_{e1}, t_{e2}]\), \([t_{a1}, t_{a2}]\), \([t_{l1}, t_{l2}]\), and \([t_{o1}, t_{o2}]\) respectively. To determine the four possible starting times for the current employee \(i\), the Sessions Scheduler randomly selects a value from each of the four corresponding intervals. Thus, for employee \(i\), there exist four distinct candidate starting times denoted as \(t_{\text{start\_abnormal\_early}}\), \(t_{\text{start\_abnormal\_late}}\), \(t_{\text{start\_late}}\), and \(t_{\text{start\_on\_time}}\). In the configuration file, operators can define different probability intervals for each possible starting time, namely \([p_{e1}, p_{e2}]\), \([p_{a1}, p_{a2}]\), \([p_{l1}, p_{l2}]\), and \([p_{o1}, p_{o2}]\). It is noteworthy that drawing a probability value from intervals \([p_{e1}, p_{e2}]\) and \([p_{a1}, p_{a2}]\) will be typically very small, reflecting the rarity of abnormally early and abnormally late starting times. Conversely, the interval \([p_{o1}, p_{o2}]\) will produce the highest probabilities, indicating the likelihood of employees starting on time. Consequently, the Sessions Scheduler assigns a random probability value to each candidate starting time, drawn from their respective probability intervals. The next step can be linked to tossing an unfair tetrahedron (a die with four faces), where each face represents a starting time option. The resulting face corresponds to the actual starting time of employee \(i\), denoted as \(t_{\text{start}}\), which constitutes the first value of the first tuple in the list \( \mathcal{S}_i \), i.e., \(t_1\). Thus, the Sessions Scheduler effectively determines the starting time for employee \(i\) using this probabilistic method, ensuring that each potential starting time is considered.

Selecting the end time \( t_{\text{end}} \) for each employee \( i \), denoted as the second time in the last tuple of the list \( \mathcal{S}_i \) (i.e., last logout time \( t_{2k_i} \)), undergoes a process akin to determining the start time \( t_{\text{start}} \). Similarly, the Sessions Scheduler employs a probabilistic approach, mirroring the methodology used for selecting \( t_{\text{start}} \). Dataset creators define intervals representing various end times, such as abnormally early, abnormally late, late, and on time, each associated with corresponding probability intervals.

\begin{figure}[h]
    \centering
    \includegraphics[width=0.5\textwidth]{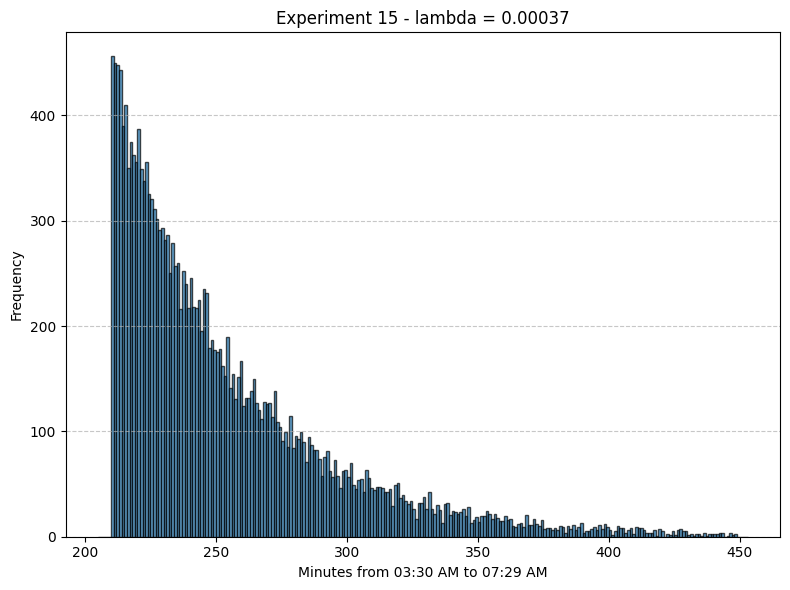}
    \caption{Frequency distribution of \( t_{\text{start\_abnormal\_early}} \) from 03:30 AM to 07:29 AM over 20,000 trials. The distribution follows an exponential distribution with a rate parameter \(\lambda = 0.00037\), indicating higher frequencies of abnormal early start times occurring at earlier minutes and tapering off towards later minutes.}
    \label{fig:start_abnormal_early}
\end{figure}

The Sessions Scheduler is not limited to drawing values from the defined time and probability intervals using a uniform distribution; it can also utilize exponential and normal distributions. For instance, consider Figure \ref{fig:start_abnormal_early}, which illustrates the Sessions Scheduler's process of selecting the value for \( t_{\text{start\_abnormal\_early}} \) over 20,000 iterations. In this example, the Sessions Scheduler is configured to draw a time value \( t \) within the interval [3:30 AM - 7:29 AM] according to an exponential distribution with a lambda \(\lambda=0.00037\), where $\lambda$ is the distribution parameter.

After defining \( t_{start} \) and \( t_{end} \), the Sessions Scheduler will determine whether employee \( i \) will have a lunch break using a similar probabilistic approach. If a lunch break is scheduled, the controller will then specify \( t_{lunch\_start} \) and \( t_{lunch\_end} \), which denote the start and end times of the lunch break, respectively.

The Sessions Scheduler manages random logouts and logins between \( t_{start} \) and \( t_{lunch\_start} \), and between \( t_{lunch\_end} \) and \( t_{end} \). This process occurs in two stages: first, the number and duration of logouts are randomly selected based on predefined intervals set in the configuration file (see Figure \ref{fig:BDE}). Next, an algorithm recursively places these logouts on the timeline while adhering to minimum and maximum gap constraints between consecutive logouts also defined in the configuration file.

\subsubsection{\textbf{Sessions Executor}}
\label{ssubsec:sessions_exec}

The Sessions Executor, a component of the Benign Data Engine (Figure \ref{fig:BDE}), manages session execution by creating a thread for each employee \(i\) in the list \( \mathcal{T} \). Using employee credentials, it runs the corresponding behavioral script \(BS_j\) on the designated host \(H_r\) at the scheduled session times in \(S_i\).

For each tuple \((t_{2j-1}, t_{2j}) \in S_i\), the Sessions Executor starts a remote session on host \(H_r\) at time \(t_{2j-1}\), runs \(BS_j\) until \(t_{2j}\), and then terminates the session. This process repeats for all tuples in \(S_i\) until the final session \((t_{2k_i-1}, t_{2k_i})\).

Each department in the system can have a general behavioral script representing typical employee actions, or individual scripts can be assigned to users, configurable in the configuration file. During each execution block (Figure \ref{fig:BDE}), a subset of behaviors from the script is randomly executed using a probabilistic approach. These behaviors include actions like browsing, downloading, running local programs, and accessing internal servers. The framework’s scripts can be customized for specific enterprise use cases, enabling the simulation of various operational environments and realistic user behaviors.

This approach provides a flexible, scalable, and adaptable method for simulating user behavior in network environments, suitable for both small and large organizations. The source code for BDE is available on our GitHub \cite{LMDG}, offering researchers access for dataset generation and other academic purposes.

\subsection{Attack Engine (AE)}
\label{subsec:attack_engine}

In this context, an \textit{Attack Engine} refers to a method or framework that enables the automated execution of cyberattacks. For example, automating DDoS attacks can often be achieved by deploying specific scripts on the attacking hosts to initiate the attack. However, as we will discuss in this section, automating LM   attacks presents unique challenges that are more complex and less straightforward than those associated with simpler scripted attacks.\\

\subsubsection{LM   Attacks}
\label{ssubsec:lateral_movement_attacks}

According to the MITRE ATT\&CK framework \cite{mitre}, nine tactics qualify as LM   techniques. These include \textit{Exploitation of Remote Services}, \textit{Internal Spearphishing}, \textit{Lateral Tool Transfer}, \textit{Remote Service Session Hijacking}, \textit{Remote Services}, \textit{Replication through Removable Media}, \textit{Software Deployment Tools}, \textit{Taint Shared Content}, and the \textit{Use of Alternate Authentication Material}. In the LMDG dataset, multiple LM   tactics from this list—such as Exploitation of Remote Services and the Use of Alternate Authentication Material—are employed, as discussed further below \ref{ssubsec:lm_attacks_in_lmdg}.

Each of these LM   tactics encompasses various techniques. For instance, the "Use of Alternate Authentication Material" tactic can be executed through techniques like "Pass-the-Hash" or "Pass-the-Ticket" attacks. To clarify the complexity and unique nature of LM   attacks compared to more straightforward attack types, we provide a detailed example of one of these attacks. This analysis highlights the operational challenges and automation complexities inherent in implementing these advanced tactics.

\begin{figure}[h]
    \centering
    \includegraphics[width=0.5\textwidth]{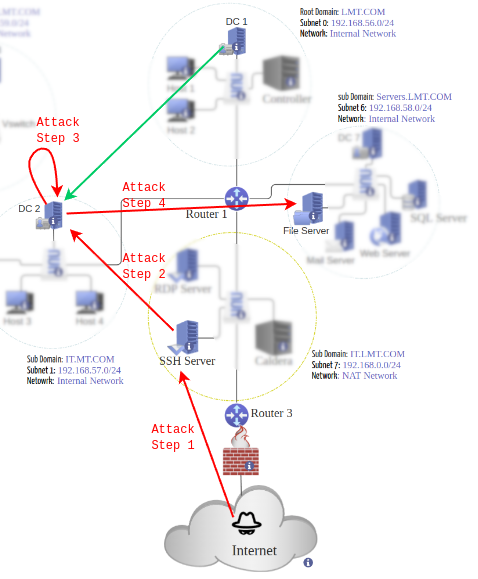}
    \caption{First Attack Scenario in LMDG dataset which is Passing the Hash attack (PtH), providing a step-by-step visualization of the movement through network nodes.}
    \label{fig:attack1}
\end{figure}

One of the attack scenarios demonstrated in the LMDG dataset involves a \textit{pass-the-hash (PtH)} attack. An outline of the attack sequence is depicted in Figure \ref{fig:attack1}. The scenario begins by assuming an attacker has obtained the local administrator credentials for domain controller DC2 in subnet 1, potentially through techniques like phishing. Using these credentials, the attacker initiates an SSH connection to an SSH server in subnet 7 (attack step 1) and subsequently connects to DC2 (attack step 2) via SSH using the same credentials. Once on DC2, the attacker downloads and executes \textit{Mimikatz} to extract credential hashes from the LSASS process, including those from recent sessions. In this case, an enterprise administrator recently accessed DC2 (shown by the green arrow in Figure \ref{fig:attack1}), allowing the attacker to retrieve the administrator’s credentials. The attacker gains an elevated shell with the enterprise admin hash (attack step 3), enabling access to restricted directories on a file server in subnet 6 (attack step 4). This elevated access allows sensitive information to be exfiltrated from a folder accessible only to the enterprise administrator.

\subsubsection{Challenges in Automating LM   Attacks}
\label{ssubsec:challenges_in_automating_lm_attacks}

As shown in the earlier attack example \ref{ssubsec:lateral_movement_attacks}, elevated and reverse shells are common in LM attacks. Automating such steps introduces key challenges. For instance, automating the \textit{Pass-the-Hash} (PtH) attack from subsection \ref{ssubsec:lateral_movement_attacks} and Figure \ref{fig:attack1} involves feasible steps like SSH-ing with stolen credentials and executing commands on the domain controller (e.g., running Mimikatz). However, automating PtH to obtain an elevated shell is difficult, as the resulting \textit{cmd.exe} process runs under elevated credentials and is hard to access without knowing its properties (e.g., PID). Reverse shells add further complexity, especially when spawned on different hosts. If the NTLM hash must also be dynamically extracted mid-attack, automation becomes even harder. These issues highlight two main automation challenges: managing elevated/reverse shells and dynamically retrieving and reusing data from earlier attack steps.

\subsubsection{A Candidate Solution}
\label{ssubsec:candidate_solution}

A solution to automating elevated and reverse shells in LM attacks, as discussed in section \ref{ssubsec:challenges_in_automating_lm_attacks}, is a client-server architecture,i.e., an orchestrator issuing commands to agents deployed on hosts. 

In the attack scenario from section \ref{ssubsec:lateral_movement_attacks}, an orchestrator in subnet 7 controls agent \( A_1 \) on the SSH server. Instead of SSH-ing from \( A_1 \) to DC2, a new agent \( A_2 \) is spawned on DC2, executing commands like running \textit{Mimikatz}. For the PtH operation, the PtH command’s \texttt{/run} parameter is modified to spawn an elevated agent \( A_3 \), completing the final attack step to exfiltrate data.

\begin{figure}[h]
    \centering
    \includegraphics[width=0.5\textwidth]{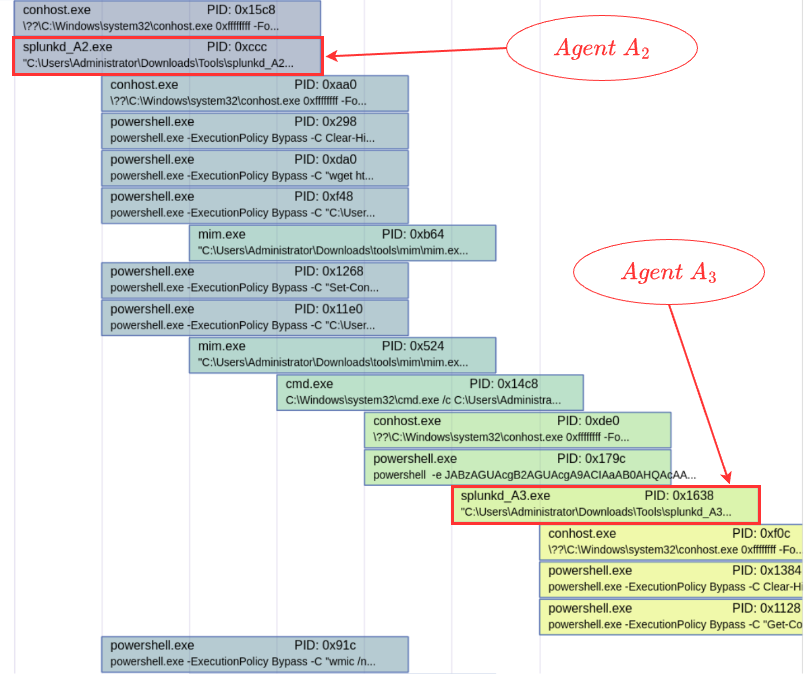}
    \caption{Partial process tree illustrating the execution of the Pass-the-Hash (PtH) attack on DC2, as discussed in subsection \ref{ssubsec:lateral_movement_attacks}. }
    \label{fig:pth_process_tree}
\end{figure}

Figure \ref{fig:pth_process_tree} shows the process tree during a Pass-the-Hash (PtH) attack using a client-server architecture. Agent $A_2$, represented by \texttt{splunkd\_A2.exe}, is deployed on Domain Controller DC2 and performs the PtH attack. It then spawns Agent $A_3$ (\texttt{splunkd\_A3.exe}) with elevated privileges. Agent $A_3$ communicates with the controller to continue the attack, accessing a restricted directory to retrieve data. This hierarchical structure visualizes how the client-server model manages privilege escalation, with attack steps chaining through communication between spawned agents. Tools like \textit{CALDERA} support this feature \cite{lademu}.

\subsubsection{CALDERA as an Attack Engine}
\label{ssubsec:caldera_as_an_attack_engine}

Following the approach outlined in \cite{lademu}, we utilized Caldera to implement the client-server architecture discussed in subsection \ref{ssubsec:candidate_solution} to automate attack execution steps. As previously noted, Caldera, along with other tools referenced in \cite{lademu}, can be employed to facilitate this level of attack automation. We refer to such tools collectively as "attack engines."

Caldera\textsuperscript{\texttrademark} \cite{caldera} is an adversary emulation platform developed by MITRE for autonomous breach-and-attack simulations, manual red-team operations, and automated incident response. Based on the MITRE ATT\&CK\textsuperscript{\texttrademark} \cite{mitre} framework, Caldera includes a core system consisting of the main framework code, an asynchronous command-and-control (C2) server, a REST API, and a web interface. It also supports plugins—separate repositories that extend the core functionality by adding agents, graphical interfaces, and collections of Tactics, Techniques, and Procedures (TTPs), enabling a flexible and comprehensive approach to adversary emulation.

\subsubsection{LM Attacks in LMDG Dataset}
\label{ssubsec:lm_attacks_in_lmdg}

The LMDG dataset contains seven attack scenarios that achieve LM   using various tactics and techniques (see table \ref{tab:lm_attacks}). Of these, three attacks were successful, while four were unsuccessful. We discuss possible reasons for each unsuccessful attack, considering that our setup includes Windows 10, Windows 11, and Windows Server 2022—the latest Windows versions with advanced security mechanisms \cite{LMDG}. This combination of successful and unsuccessful attacks is valuable for understanding attacker behavior, as many attacks tend to fail due to robust defenses, with only some achieving success.

Our dataset includes multiple versions of each attack, targeting different hosts and subnets. In some cases, attacks were executed repeatedly to enrich the dataset with diverse instances of attack records; we refer to this repetition of the same scenario, version pair, as a trial.

The attack steps depicted in the figures \cite{LMDG}, e.g., figure \ref{fig:attack1}, within the attack explanations represent LM   hops, as defined in Section \ref{sec:discussion}. All attack scenarios share the first two steps: initial access to the SSH server from outside the network using stolen credentials, followed by access to an additional internal machine. Beyond these initial steps, each attack scenario diverges in tactics and execution. More details about attacks execution are presented in \ref{sec:dataset} and our Github \cite{LMDG}.

\begin{table*}[ht!]
\centering
\begin{tabular}{|c|p{15cm}|}
\hline
\textbf{Attack} & \textbf{Description} \\ \hline
\textbf{Passing the Hash} & Use the hashed password of an enterprise admin to authenticate and gain unauthorized access. \\ \hline
\textbf{Asreproastable} & Exploit a user's AS-REP response to steal credentials and impersonate the administrator to steal data. \\ \hline
\textbf{Pass the TGT} & Dump LSASS memory to identify a domain admin, steal the TGT, inject it into memory, and steal data. \\ \hline
\textbf{Attack Delegation} & Perform AS-REP roasting, steal credentials, abuse group permissions (e.g., AddSelf), execute DC Sync to steal administration credentials, renew the TGT, and perform actions as the administrator. \\ \hline
\textbf{Password Spray} & Use password brute-forcing to open a zip file, perform a password spray attack to log in as an admin, abuse write permissions on a specific share to add a malicious file executed by a domain admin, and steal data. \\ \hline
\textbf{Silver Ticket} & Dump LSASS memory to identify a domain admin, use the stolen data to create a silver ticket, inject it into memory, and perform actions as the domain admin. \\ \hline
\textbf{Golden Ticket} & Dump LSASS memory to identify a domain admin, use the stolen data to create a golden ticket, inject it into memory, and perform actions as the domain admin. \\ \hline
\end{tabular}
\caption{Summary of Lateral Movement Attacks in LMDG Dataset}
\label{tab:lm_attacks}
\end{table*}

\subsection{LMDG Labelling Engine (LE)}
\label{subsec:labelling_engine}

In cybersecurity datasets, labeling involves identifying and extracting records associated with attack activities from system logs and network traffic. The Labeling Engine serves as the component responsible for automating this extraction process. This subsection will examine the challenges of achieving accurate labeling and introduce our innovative labeling methodology.

\subsubsection{Challenges in Attack Data Labeling}
\label{ssubsec:challenges_in_attack_data_labeling}

A review of labeling techniques for cybersecurity dataset generation reveals three primary approaches: \textbf{\textit{Injection Timing}}, \textbf{\textit{Behavioral Profiling}}, and \textbf{\textit{Network Security Tools}} \cite{Human-guidedAuto-LabelingForNetworkTrafficData:TheGELMapproach}.

The \textbf{\textit{Injection Timing}} approach labels all logs or network traffic within the attack period as malicious, improving accuracy when combined with other methods \cite{landauer2020, TowardsGeneratingRealLifeDatasetsForNetworkIntrusionDetection, creech2013generation, garcia2014empirical, haider2017generating}. However, it assumes no benign events occur during the attack, which leads to inaccuracies, particularly in complex attacks like LM, where benign activity may overlap with malicious actions.

The \textbf{\textit{Behavioral Profiles}} method uses predefined profiles of malicious and benign behaviors for labeling \cite{creme, shiravi2011survey, ring2017flow}. While effective for identifying attack-specific characteristics, it fails in LM attacks, where legitimate users and hosts are exploited, rendering behavioral profiling insufficient.

The \textbf{\textit{Network Security Tools}} approach uses data from security tools such as IDS, honeypots, and packet sniffers to label records \cite{aparicio2014automatic, ring2017flow}. Despite its utility, this method faces accuracy issues, including false positives and negatives, due to tool limitations.

Given these challenges, there is a need for a more accurate labeling technique. We propose a novel methodology that addresses the shortcomings of existing approaches, improving accuracy, particularly for LM and advanced persistent threats (APTs), where traditional methods are less effective.

\subsubsection{LMDG Labeling Engine}
\label{ssubsec:lmdg_labeling_engine}

Our labeling methodology builds upon and extends the labeling approach introduced in \cite{lademu}, with specific enhancements and improvements outlined in the related work section \ref{paragraph:lademu_framework}. We designate this approach as \textbf{\textit{process tree labeling}}, which can be considered an additional automatic labeling technique and, as we argue, the most accurate among those reviewed. The effectiveness of process tree labeling relies on the client-server architecture introduced in \ref{ssubsec:candidate_solution} and \ref{ssubsec:caldera_as_an_attack_engine} for automating attack execution, a dependency explored in greater detail in \cite{lademu}. 

Upon the completion of attack execution, the LMDG labeling engine, along with its input—a descriptive file containing metadata on attack steps—operates from the controller, depicted in Figure \ref{fig:network_topology}. The engine distributively performs the labeling task across each affected host, using the defined attack steps from the input file to extract the relevant subset of system logs and network connections associated with each attack stage on every affected host. The LMDG labeling engine completes this process in three primary phases: \textbf{\textit{Attack Steps Forest Construction}}, \textbf{\textit{System Logs Labeling}}, and \textbf{\textit{Network Traffic Labeling}}. Before detailing these stages, we will first discuss the input to this engine, namely the descriptive file containing attack steps metadata.\\

\paragraph{\textbf{\textit{LMDG Labeling Engine Input}}}
\label{paragraph:lmdg_labeling_engine_input}

The input to the labeling engine consists of a set of hosts impacted by various attack steps, where each host includes a collection of malicious processes with specific attributes. Let $H$ represent the set of all such hosts, i.e., $H = \{h_1, h_2, ..., h_n\}$, with each host $h_i \in H$ uniquely identified by a $\mathtt{HostName}$. For each host $h_i$, let $P(h_i)$ denote the set of processes associated with the malicious agents deployed on that host during any attack step, i.e., $P(h_i) = \{p_1, p_2, ..., p_m\}$. Each process $p_j \in P(h_i)$ is described by the following tuple

\begin{align*}
p_j = (&\pi, \; t_s, \; t_e, \; \sigma, \; \nu, \; \tau, \; \kappa, \; \phi)
\end{align*}

In this tuple, $\mathtt{\pi}$ denotes the process identifier associated with a deployed Caldera agent, i.e., PID. $t_s$ and $t_e$ define the time window during which a particular attack step occurred (start time and end time). The specific step within the attack and the overarching scenario are identified by the $\kappa$ and $\sigma$ fields in the tuple, with $\phi$ indicating whether the step was completed successfully. Since an attack scenario can be executed across various hosts or subnets, multiple versions of the same scenario may exist. For example, in subsection \ref{ssubsec:lateral_movement_attacks}, the pass-the-hash (PtH) attack can be executed on different subnets (e.g., subnet 4 instead of subnet 1). This versioning is captured by the $\nu$ field in the tuple. Additionally, we may execute the same scenario version multiple times; thus, the $\tau$ field is included to distinguish between these instances, offering clear differentiation across repeated executions or trials.

Each host \( h_i \in H \) can be formally represented as a tuple containing its \( \mathtt{HostName} \) and the set of associated processes, \( P(h_i) \), for each Caldera agent deployed on that host. Formally, this is expressed as \( h_i = (\mathtt{HostName}_i, P(h_i)) \). The overall input structure can then be denoted by \( \text{Input} = \{ h \mid h \in H \} \). This structured organization facilitates the grouping of processes by host, enabling efficient distribution of the labeling process and correlation of process executions with the various stages of distinct attack scenarios.

\paragraph{\textbf{\textit{Attack Steps Forest
Construction}}}
\label{paragraph:attack_steps_forest_construction}

The first stage in our labeling engine is the \textit{Attack Steps Forest Construction}, where we build a forest \( \mathcal{F} \) of \( m \) process trees \( \mathcal{T}_{p_j} \), one for each malicious process \( p_j \in P(h_i) = \{p_1, \ldots, p_m\} \) on host \( h_i \). Each tree is rooted at \( p_j \)'s PID and includes all descendant processes within its execution window \([t_s, t_e]\). This temporal constraint ensures that each \( \mathcal{T}_{p_j} \) captures causally relevant attack activity, providing the foundation for accurate step-level labeling.

\begin{algorithm}[H]
\caption{Attack Steps Forest Construction}
\label{alg:attack_steps_forest}
\begin{algorithmic}[1]
\Procedure{AttackStepsForestConstruction}{$H$} 
    \State $\mathcal{F} \gets \emptyset$ \Comment{Initialize the forest of attack steps}
    
    \For{$h_i \in H$} 
        \For{$p_j \in P(h_i)$}
            \State $\pi \gets p_j.\pi$, \; $t_s \gets p_j.t_s$, \; $t_e \gets p_j.t_e$
            \State $\sigma \gets p_j.\sigma$, \; $\nu \gets p_j.\nu$, \; $\tau \gets p_j.\tau$
            \State $\kappa \gets p_j.\kappa$, \; $\phi \gets p_j.\phi$ \Comment{Extract $p_j$'s attributes}
            \State $\mathcal{L} \gets \emptyset$ \Comment{Initialize a list for process IDs}
            
            \State $\mathcal{T}_{p_j} \gets$ \Call{GetProcessTree}{$\pi, t_s, t_e, \mathcal{L}$} \Comment{Build process tree of process $p_j$}
            
            \State $\mathcal{T}_{{p_j}_{\text{meta}}} \gets (\mathcal{T}_{p_j}, t_s, t_e, \sigma, \nu, \tau, \kappa, \phi)$
            \State $\mathcal{F} \gets \mathcal{F} \cup \{\mathcal{T}_{{p_j}_{\text{meta}}}\}$
        \EndFor
    \EndFor
    \State \Return $\mathcal{F}$
\EndProcedure

\Procedure{GetProcessTree}{$\pi, t_s, t_e, \mathcal{L}$}
    \State $\mathcal{L} \gets \mathcal{L} \cup \{\pi\}$
    \State $\mathcal{E} \gets \{ e \in \mathcal{E}_{h_i}^{4688} \; | \; e.\pi = \pi \; \wedge \; t_s \leq e.t \leq t_e \}$
    \For{$e \in \mathcal{E}$}
        \If{$e.\pi \notin \mathcal{L}$}
            \State $\mathcal{L} \gets \mathcal{L} \cup \{e.\pi\}$
            \State \Call{GetProcessTree}{$e.\pi, t_s, t_e, \mathcal{L}$}
        \EndIf
    \EndFor
    \State \Return $\mathcal{L}$
\EndProcedure

\end{algorithmic}
\end{algorithm}

The specifics of this step are outlined in Algorithm \ref{alg:attack_steps_forest}, where the set \( \mathcal{E}_{h_i}^{4688} \) represents all process creation events recorded in the Windows Security log for host \(h_i\).

\begin{figure}[h]
    \centering
    \includegraphics[width=0.5\textwidth]{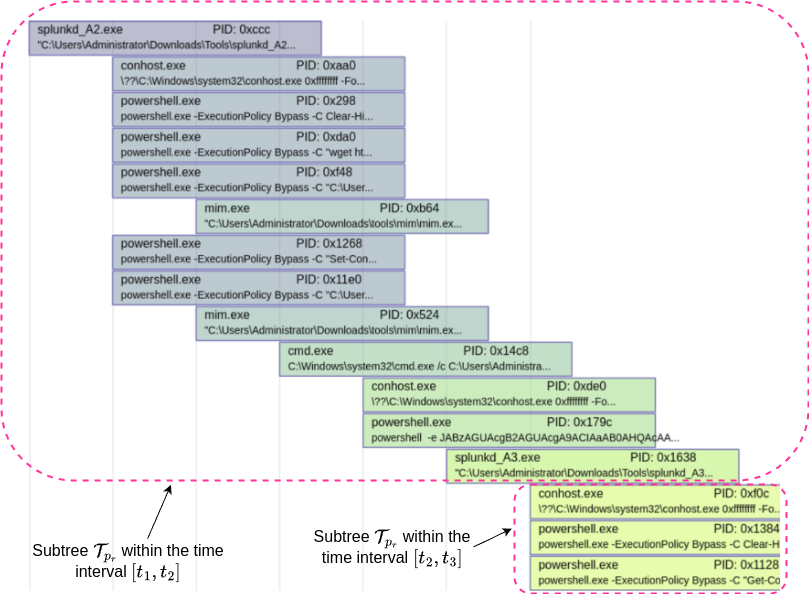}
    \caption{Output of algorithm \ref{alg:attack_steps_forest} showing two process trees rooted at the same malicious process \(p_r\). The first tree, representing attack step 3, and the second tree, representing attack step 4 in Figure \ref{fig:attack1} explained in subsection \ref{ssubsec:lateral_movement_attacks}.}
    \label{fig:asfc_process_tree}
\end{figure}

An example of the output generated by Algorithm \ref{alg:attack_steps_forest} is shown in Figure \ref{fig:asfc_process_tree}. This output corresponds to the example previously detailed in Subsection \ref{ssubsec:lateral_movement_attacks}, which illustrates a Pass-the-Hash (PtH) attack scenario, as depicted in Figure \ref{fig:attack1}. 

In Figure \ref{fig:asfc_process_tree}, the example demonstrates two process trees rooted at the same process ID \(\pi\) of \(p_r\). The first tree is constructed within the constrained time interval \([t_1, t_2]\), encompassing all subprocesses that occurred within this interval and represents attack step 3 from Figure \ref{fig:attack1}. The second tree is built under the time constraint \([t_3, t_4]\), including all subprocesses within this later interval, and represents attack step 4 from the same figure \ref{fig:attack1}.

We next move to the subsequent steps, System Logs Labeling and Network Traffic Labeling, which depend on the constructed attack steps forest \(\mathcal{F}\).

\paragraph{\textbf{\textit{System Logs Labeling}}}
\label{paragraph:system_logs_labeling}

In this step, for each host \( h_i \), we iterate over the set \( \mathcal{L}_{h_i} \), which represents the collection of all Windows event logs on that host. For each log \( l \in \mathcal{L}_{h_i} \), we further iterate over the trees in the constructed forest \( \mathcal{F} \) at host \( h_i \), where each tree \( \mathcal{T}_{p_j} \in \mathcal{F} \) corresponds to a specific attack step. The primary objective here is to examine whether the current log \( l \) contains any events with process IDs matching those within the current tree \( \mathcal{T}_{p_j} \) that occurred within the specified time interval \([t_s, t_e]\). If such events are found, they are extracted and tagged with metadata of the current tree \( \mathcal{T}_{p_j} \), including details like the attack scenario, version, step number, and step success status. This labeling process operates at the \textit{\textbf{attack step level}}, incorporating relevant MITRE ATT\&CK tactics and techniques to contextualize each event within the broader attack framework.

\paragraph{\textbf{\textit{Network Traffic Labeling.}}}
\label{paragraph:network_traffic_labeling}

In this step, we construct the set \( \mathcal{E}_{h_i}^{5156} = \{ e \mid e.\text{EventID} = 5156 \wedge e \in \mathcal{E}_{h_i} \} \), which represents the collection of Windows events with Event ID 5156, corresponding to the Windows Filtering Platform (WFP). The WFP monitors and filters network traffic on Windows systems, and \( \mathcal{E}_{h_i} \) denotes the set of all Windows event logs at host \( h_i \). Subsequently, we iterate over the trees in the constructed forest \( \mathcal{F} \) and examine whether any process ID in the current tree \( \mathcal{T}_{p_j} \) matches a process ID from the events in \( \mathcal{E}_{h_i}^{5156} \). If a match is found, we filter the relevant events and label them with the metadata of the corresponding tree \( \mathcal{T}_{p_j} \), including attack scenario, version, attack step, step success, and so on. Similar to the System Logs Labeling step, we also consider MITRE ATT\&CK tactics and techniques for contextualizing the events within the attack framework.

The process of network flow labeling can be effectively performed using the packet capture (PCAP) files collected from each host, as mentioned in subsection \ref{subsec:testbed} and the output of this labeling step.

Our "Process Tree Labeling" methodology provides superior accuracy in automated labeling by associating process activities with specific attack steps. By utilizing temporal and contextual information from system logs and network traffic, it ensures precise event attribution within the attack lifecycle, enhancing labeling fidelity and reliability.

\section{Dataset}
\label{sec:dataset}

The experimental environment consists of 25 VMs, including a Controller, Caldera server, domain controllers, application servers, hosts, and routers. While 22 user accounts were set up, only 11 credentials were used by the Benign Data Engine to generate benign data. Windows Event logs and PCAP files were collected from all Windows machines except the Controller and Caldera server. PCAP files were also captured from routers 1 and 2 for supplementary network data.

The dataset was generated over 25 days (October 10–November 3, 2024), with continuous benign data generation by the Benign Data Engine. Attacks occurred between October 23 and November 1, 2024, resulting in both benign and malicious data during this period. The dataset contains only benign data before October 23, 2024.

\begin{figure}[h]
    \centering
    \includegraphics[width=0.5\textwidth]{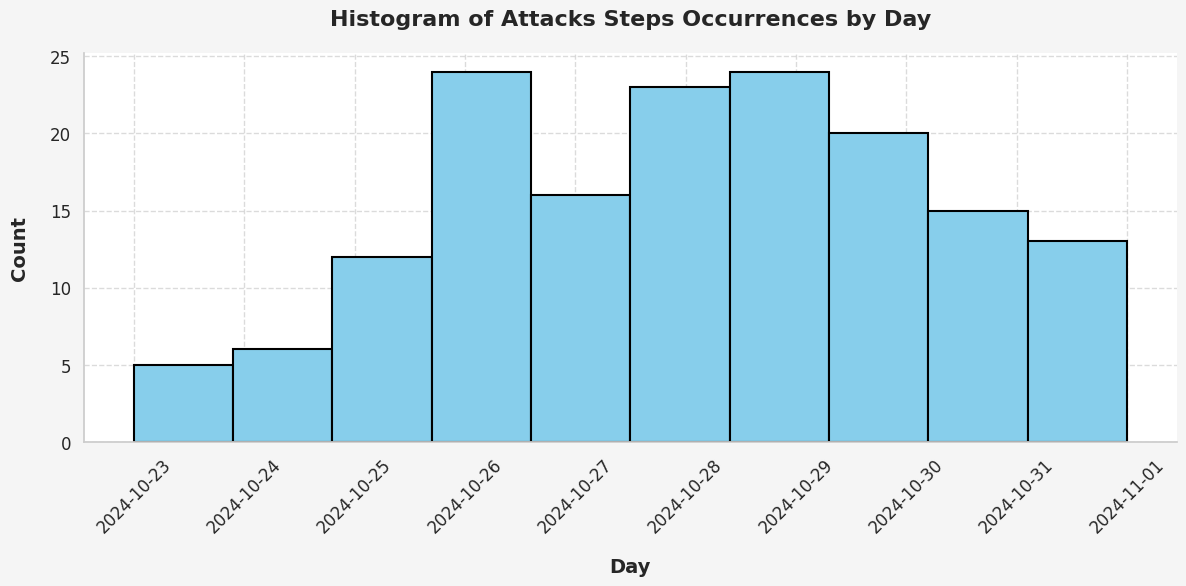}
    \caption{\textbf{Daily Distribution of Attack Steps}: This histogram visualizes the frequency of attack steps executed over time, with each bar representing the count of attack steps occurring on a specific day. The x-axis denotes individual days, while the y-axis represents the number of occurrences.}
    \label{fig:attacks_steps_histogram}
\end{figure}

We present statistics on attacks within the LMDG dataset. Figure \ref{fig:attacks_steps_histogram} shows the \textit{Daily Distribution of Attack Steps}, illustrating the frequency of attack steps over time, with the total attack size in the dataset under 1\%.

\begin{figure}[h]
    \centering
    \includegraphics[width=0.5\textwidth]{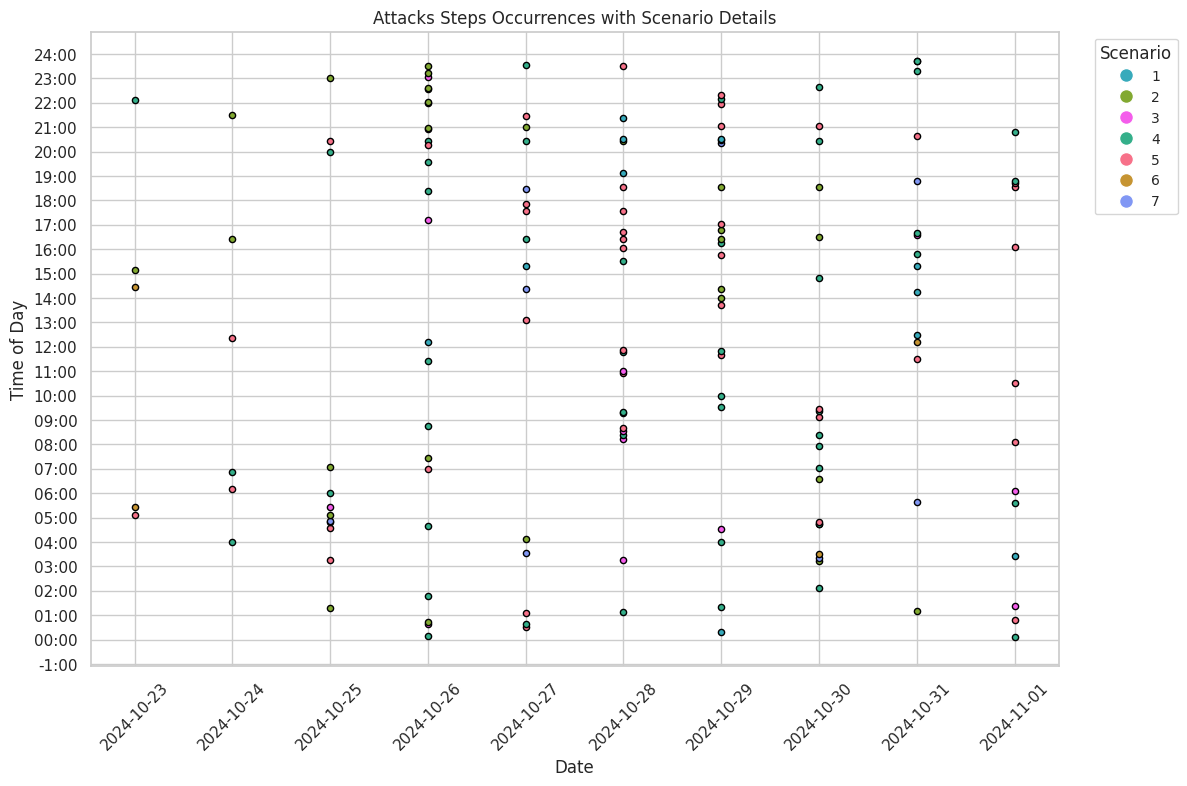}
    \caption{\textbf{Timeline of Attack Step Occurrences by Scenario}: Thacross various days, with each point representing the occurrence of an attack step on a specific day and time. The x-axis indicates the occurrence dates, while the y-axis represents the time of day to highlight daily distribution patterns. Each scenario is color-coded with a distinct hue, allowing for quick differentiation of scenarios.}
    \label{fig:attacks_scatter_plot}
\end{figure}

Figure \ref{fig:attacks_scatter_plot} displays the \textit{Timeline of Attack Step Occurrences by Scenario}, showing attack step timings via a scatter plot with distinct color coding for different scenarios.

\begin{figure}[h]
    \centering
    \includegraphics[width=0.5\textwidth]{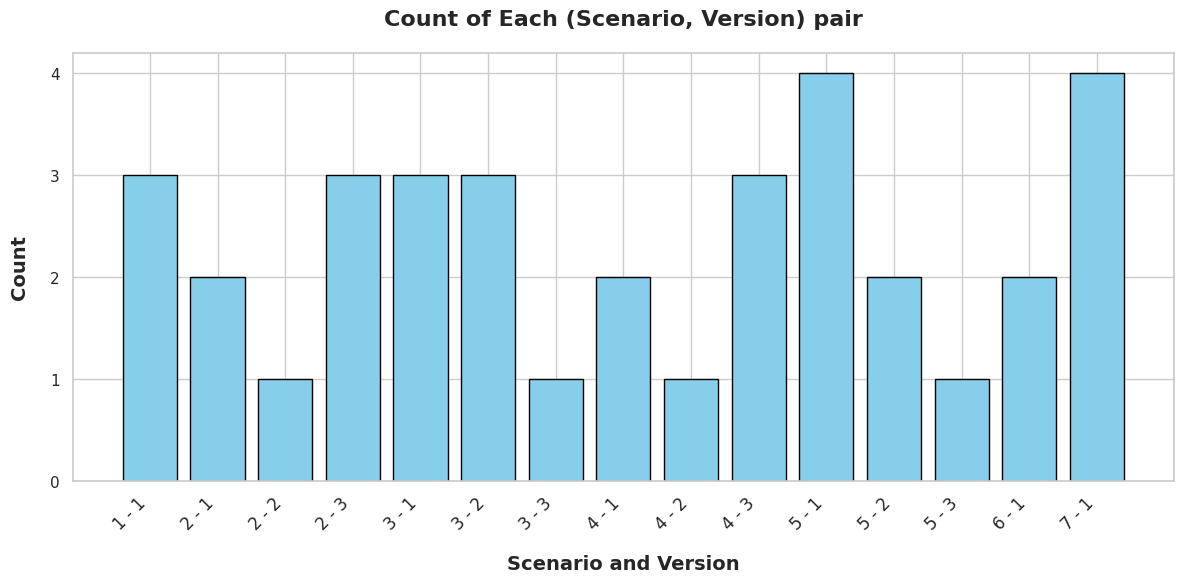}
    \caption{\textbf{Frequency Distribution of Scenario and Version pairs}: This bar plot displays the count of occurrences for each distinct (Scenario, Version) pair. The x-axis represents individual combinations of scenarios and their respective versions. The y-axis shows the count of occurrences.}is scatter plot illustrates the timing of attack steps 
    \label{fig:scenario_version_histogram}
\end{figure}

Figure \ref{fig:scenario_version_histogram} presents the \textit{Frequency Distribution of Scenario and Version Pairs}, comparing attack frequencies across scenario versions. The full execution timeline is available on our GitHub repository \cite{LMDG}.

The compressed dataset size is 253 GB (excluding router data) and 527 GB (including router data), with router 1's PCAP at 201 GB and router 2's at 72 GB. The total uncompressed dataset is 944 GB, comprising 900.93 GB of PCAP files and 43.38 GB of system logs (Table \ref{table:dataset_statistics}).

Like the LANL 2015 dataset, this dataset enables detailed extraction of authentication features and patterns, supporting research in intrusion detection, behavioral analysis, and user activity monitoring. Its rich event data and metadata provide a solid foundation for diverse cybersecurity research tasks.

\begin{table}[h!]
    \centering
    \caption{Dataset Statistics}
    \begin{tabular}{lcc}
        \toprule
        \textbf{Statistic} & \textbf{PCAP Size (GB)} & \textbf{Log Size (GB)} \\
        \midrule
        Total Size & 900.93 & 43.38 \\
        Average Size & 37.54 & 2.17 \\
        Minimum Size & 0.51 & 0.41 \\
        Maximum Size & 451.00 & 4.97 \\
        %Median Size & 6.39 & 1.41 \\
        %Standard Deviation & 95.99 & 1.40 \\
        \bottomrule
    \end{tabular}
    \label{table:dataset_statistics}
\end{table}

\section{Qualitative Analysis}
\label{sec:qualitative_analysis}

Dataset quality depends on several phases: testbed design, benign data generation, log collection, attack execution, and labeling. Enhancing these phases improves dataset robustness.

Our testbed (Section \ref{subsec:testbed}) simulates a realistic enterprise network using virtualization for scalable and detailed emulation. The Benign Data Engine (BDE) (Section \ref{ssubsec:lmdg_labeling_engine}, Figure \ref{fig:BDE}) generates realistic benign data based on behavioral scripts. These scripts, modeled on departmental roles with randomization, replicate typical employee activities such as logins, web browsing, service requests, and program execution.

Our dataset includes comprehensive system logs and network traffic data, with labeling applied to both, ensuring complete activity records for detailed behavior analysis.

Existing LM datasets have limitations such as few LM instances, outdated patterns, limited techniques, short timeframes, and minimal hops. In contrast, our LM attacks (Section \ref{ssubsec:lm_attacks_in_lmdg}) address these issues with diverse techniques, recent patterns, a 10-day execution window, and up to 7 hops across hosts, users, and subnets. Tools like CALDERA further improve attack design flexibility.

Our automated \textit{process tree labeling} (Section \ref{subsec:labelling_engine}) achieves superior accuracy, effectively tracking process hierarchies critical for LM and APT attacks.

The LMDG framework is designed to support the generation of high-quality datasets through its various integrated components. The LMDG dataset serves as an exemplar of this capability, demonstrating the framework's effectiveness in producing datasets that are comprehensive, well-structured, and suitable for advanced research and analysis.

\section{Related Work}
\label{sec:related_work}

\subsection{LADEMU Framework}
\label{paragraph:lademu_framework}

LADEMU ~\cite{lademu} uses virtualization tools (e.g., VirtualBox, VMware) for \textbf{\textit{testbed infrastructure}}, similar to LMDG. For \textbf{\textit{dataset collection}}, it captures network traffic (pcap) and Sysmon log, whereas LMDG expands coverage by collecting all Windows event logs and labeling them comprehensively. In \textbf{\textit{benign data generation}}, LADEMU employs GHOST, while LMDG offers greater flexibility through its Benign Data Engine (BDE) (\ref{subsec:BDE}), which separates session scheduling from behavior execution. Both frameworks use Caldera for \textbf{\textit{attack execution}}~\ref{subsec:attack_engine}. For \textbf{\textit{labeling}}, LADEMU builds malicious process trees from Sysmon logs, a method LMDG also adopts \ref{subsec:labelling_engine} with four key improvements. LADEMU constructs malicious process trees rooted at the Caldera agent and labels all benign process events during interaction intervals $[t_1, t_2]$ as malicious~\cite{landauer2020}. This can lead to mislabeling, especially in cases like code injection, where benign processes may act maliciously over extended periods. In contrast, LMDG labels only events linked to malicious trees, avoiding such inaccuracies. While LADEMU uses Caldera's start and finish times to bound trees~\cite{landauer2020}, this can miss delayed malicious activity. LMDG extends the end time beyond Caldera’s to ensure full coverage. LADEMU includes C\&C signals in logs~\cite{landauer2020}, reducing realism. LMDG filters these out to better mimic real-world scenarios. Finally, LMDG links labels to attack steps and scenarios~\cite{lademu}, enabling more context-aware datasets and improving multi-step attack detection.

\subsection{AIT Framework}
% \paragraph{\textbf{\textit{AIT Framework}}}
\label{paragraph:ait_framework}

The AIT framework by Landauer et al.~\cite{landauer2020, landauer2022, landauer2023} offers a model-driven approach~\cite{UsingModel-drivenArchitecture} to cybersecurity dataset generation, emphasizing \textit{automated testbed creation}~\cite{landauer2020, landauer2022} and a \textit{labeling methodology}~\cite{landauer2022}. The Kyoush platform, built with Terraform, Ansible, and OpenStack~\cite{landauer2020}, enables reusable, \textbf{\textit{automated testbed deployment}} but requires significant setup effort. Their \textit{\textbf{labeling}} combines injection timing~\cite{Human-guidedAuto-LabelingForNetworkTrafficData:TheGELMapproach, DatasetsAreNotEnough:ChallengesInLabelingNetworkTraffic} with manual query-based log inspection for each attack step~\cite{landauer2022}, limiting scalability. While \textbf{\textit{attack scripts}} were automated, the framework lacks support for complex attacks like LM~\ref{subsec:attack_engine}. \textbf{\textit{Data collection}} targets Linux logs, and \textit{\textbf{benign behavior}} is simulated via a User State Machine.\\

\subsection{CREME Framework}
% \paragraph{\textbf{\textit{CREME Framework}}}
\label{paragraph:creme_framework}

CREME, introduced by Bui et al.~\cite{creme}, generates labeled intrusion detection datasets and evaluates dataset quality. It employs virtualization for \textbf{\textit{testbed infrastructure}} and \textit{\textbf{collects data}} via tcpdump, rsyslog, and Atop, but supports only Linux/Unix systems. For \textbf{\textit{benign data generation}}, its "Reproduction Module" runs unspecified benign programs; it also executes five \textit{\textbf{attack}} types, though support for complex LM attacks is unclear~\ref{subsec:attack_engine}. \textit{\textbf{Labeling}} uses Behavioral Profiles~\cite{Human-guidedAuto-LabelingForNetworkTrafficData:TheGELMapproach, DatasetsAreNotEnough:ChallengesInLabelingNetworkTraffic}, tagging all traffic from attack machines as malicious, which can not handle LM attacks~\ref{ssubsec:challenges_in_attack_data_labeling}.

\section{Discussion}
\label{sec:discussion}

% Write pros and cons of each step of framework
% talk about CALDAERA traffic and that it should be removed

Our review of LM   detection reveals a need for a comprehensive definition of LM  . While MITRE ATT\&CK \cite{mitre} defines it broadly as techniques for accessing and controlling remote systems, this vague description limits the effectiveness of detection models, highlighting the need for a more precise and actionable definition. 

We define two key types of adversary progression: \textit{horizontal progression} and \textit{vertical progression}. \textit{Horizontal progression} involves gaining independent access to multiple hosts without interdependence, which does not qualify as LM  . In contrast, \textit{vertical progression} describes interconnected access, where controlling one system enables access to others. \textbf{\textit{We define LM   as vertical progression across hosts, accounts, or privileges, where one access leads to another}}. This includes movement between hosts, accounts with elevated privileges, and privilege escalation. This refined definition is essential for creating effective detection models.

In the cloud environment, LM   follows a similar concept with modifications. \textit{Identities} (user, application, and service accounts) correspond to \textit{accounts}, requiring authentication to access resources, while \textit{permissions} or \textit{policies} align with \textit{privileges}, defining access levels. A key distinction in the cloud is the \textit{services layer}, which includes resources like AWS EC2 and S3, providing computing and storage. Thus, \textit{cloud LM   involves vertical progression across identities, permissions/policies, services, and resources}. For example, an attack detailed by Microsoft Threat Intelligence \cite{Azure_LM} involved exploiting SQL injection to access an Azure database server and using the Instance Metadata Service (IMDS) to obtain further access to cloud resources.\\

Regarding the threat model in the LMDG dataset, it emulates realistic APT-like scenarios where adversaries perform stealthy, persistent attacks over an extended period. These scenarios include initial access, privilege escalation, and multi-hop LM   across hosts and network subnets, reflecting the sophisticated behaviors of modern attackers targeting enterprise networks. By leveraging the CALDERA platform for attack emulation, the framework enables flexible attack design, supporting a variety of LM   techniques and bypassing typical security defenses. To enhance realism, these attacks occur within a backdrop of benign user activities generated by the Benign Data Engine (BDE), providing a nuanced environment for distinguishing between normal and malicious behavior. This threat model thus offers a robust foundation for evaluating detection and response mechanisms against complex and dynamic cyber threats.

\section{Conclusions and Future Work}

In this work, we have comprehensively examined current cybersecurity benchmark datasets with a specific focus on evaluating the presence and characteristics of LM     attacks. Our analysis, the first of its kind, assessed LM datasets across multiple dimensions, including the quantity and variety of LM techniques, attack duration, number of movement hops, data sources (e.g., authentication logs, network flows), labeling methodologies, and testbed configurations. This investigation has highlighted gaps and challenges within existing datasets, providing insight into the strengths and limitations of current approaches to LM   detection.

We developed a benchmark dataset focused explicitly on LM   attacks to address the identified limitations. This dataset, designed to overcome many existing issues in LM datasets, provides a valuable resource for the research community, facilitating the training and evaluation of more effective LM detection models. Our qualitative dataset analysis demonstrates its applicability for various LM   scenarios. It ensures that the diversity and complexity of attacks are suitable for testing advanced detection techniques.

Additionally, we introduced the LM   Dataset Generator (LMDG) framework, a reproducible toolset for generating high-quality LM and APT datasets. The LMDG framework automates benign data generation, attack execution, and—crucially—the labeling of attack-related events in system and network logs. Recognizing the challenges posed by automatic labeling in LM scenarios, where benign hosts may perform malicious actions, we proposed a novel technique, \textit{process tree labeling}. This method offers improved precision and accuracy over existing techniques such as injection timing, behavior profiles, and network security tools. Overall, the contributions of this work enhance the landscape of LM dataset generation and analysis, supporting further advancements in cybersecurity research and LM detection capabilities.

Several limitations of our framework warrant consideration. First, using virtualization to construct testbeds necessitates extensive domain expertise, making the process time-consuming and highly case-dependent, as discussed in more detail in \cite{landauer2020}. This requirement for specialized knowledge may hinder the framework's scalability and accessibility. Second, the client-server architecture employed in attack automation, as outlined in Sections \ref{ssubsec:candidate_solution} and \ref{ssubsec:caldera_as_an_attack_engine}, introduces traffic and log accuracy challenges. Specifically, the traffic generated by client-server communication must be filtered to avoid contaminating the dataset with automation-related signals, ensuring that the resulting data remains realistic and reflective of actual attack behaviors. Finally, our proposed labeling methodology, process tree labeling, the most accurate automatic labeling technique, is inherently tied to the client-server automation model. This dependency arises from the need to identify the process IDs of deployed agents, creating a coupling between labeling and attack automation. This coupling is discussed in more detail in \cite{lademu} and may limit the applicability of our labeling approach in environments where such client-server structures are not feasible.

While this study includes a qualitative analysis of our dataset \ref{sec:qualitative_analysis} and comparisons with existing datasets in the literature, further work is needed to incorporate quantitative analysis methods. A systematic review of current quantitative assessment techniques used in cybersecurity datasets will enable us to apply rigorous, data-driven evaluation metrics to our dataset, enhancing its reliability and usability. In addition, future efforts may focus on producing a more comprehensive dataset that encompasses the full spectrum of Advanced Persistent Threat (APT) attack stages rather than concentrating solely on LM  . Such a dataset would capture all phases of APT attacks, offering a richer resource for developing and benchmarking holistic detection models that address the complete lifecycle of sophisticated attack vectors. This extension will advance research into multi-stage threat detection, providing excellent value for the cybersecurity community.

% \clearpage
\bibliographystyle{plain}
\bibliography{references}

\end{document}